\documentclass[pra,twocolumn,amsmath,amssymb]{revtex4}

\newcommand{\beq}{\begin{equation}}
\newcommand{\eeq}{\end{equation}}
\newcommand{\beqa}{\begin{eqnarray}}
\newcommand{\eeqa}{\end{eqnarray}}
\newcommand{\ket} [1] {\vert #1 \rangle}
\newcommand{\bra} [1] {\langle #1 \vert}

\newcommand{\proj}[1]{\ket{#1}\bra{#1}}

\newcommand{\opnorm}[1]{|\!|\!|#1|\!|\!|_2}
\newcommand{\newatop}[2]{\genfrac{}{}{0pt}{}{#1}{#2}}

\begin{document}

\title{Adiabatic quantum search algorithm for structured problems}
\author{J\'er\'emie Roland}
\author{Nicolas J. Cerf}
\affiliation{Center for Quantum Information and Communication, CP 165/59,
Ecole Polytechnique, Universit\'e Libre de Bruxelles, 1050 Brussels, Belgium}

\date{\today}

\begin{abstract}
The study of quantum computation has been motivated by the hope of finding efficient quantum algorithms for solving classically hard problems.
In this context, quantum algorithms by local adiabatic evolution have been shown to solve an unstructured search problem with a quadratic speed-up over a
classical search, just as Grover's algorithm. In this paper, we study how the structure of the search problem may be exploited to further improve
the efficiency of these quantum adiabatic algorithms. We show that by nesting a partial search over a reduced set of variables into a global search, it is possible to devise quantum adiabatic algorithms with a complexity that, although still exponential, grows with a reduced order in the problem size.
\end{abstract}

\maketitle

\section{Introduction}
Grover's quantum algorithm solves an unstructured search problem in a time of order $\sqrt{N}$, where $N$ is the dimension of the search space, which corresponds to a quadratic speed-up over a classical search \cite{grov96}.
This algorithm is proved to be optimal in the case of unstructured search problems \cite{zalk99}. Naturally, it can also be used to solve a
structured search problem with a quadratic speed-up over a naive classical search that would exhaustively check every possible solution. However,
exploiting the structure of the problem is well known to lead to better classical
search algorithms. It is therefore tempting to imagine that better quantum search algorithms may be devised similarly by exploiting the problem structure. 
Following this, Cerf, Grover and Williams have shown that this could be done by partitioning the unknown variables into two (or more) sets and nesting a quantum search over one set into another search over two (or more) sets, yielding an average complexity of order $\sqrt{N^\alpha}$, with $\alpha<1$ \cite{cerf00:nesting}.

While this algorithm, as well as Grover's original algorithm, stay within the standard paradigm of quantum computation based on quantum circuits, a new type of quantum algorithms based on adiabatic evolution has been introduced lately \cite{farh00,vdam01}. In particular, a quantum adiabatic analogue of
Grover's search algorithm has been proposed in Ref.~\cite{rola02}, 
which works for unstructured search problems. The use of quantum 
adiabatic algorithms has also been analyzed for solving structured
problems such as $k$-SAT, but in such a way that until now only a numerical study has been possible \cite{farh01}.
Recently, the study of quantum adiabatic algorithms has progressed even further after Aharonov and Ta-Shma demonstrated their universality,
namely that any problem that may be solved efficiently in the standard (circuit-based) model of quantum computation can also
be solved efficiently by a quantum adiabatic algorithm \cite{ahar03}.
This provides a strong incentive to the search for new quantum adiabatic
algorithms.

The purpose of this paper is to bring the ideas of nested quantum search and quantum adiabatic computation together, in order to devise
a new quantum adiabatic algorithm adapted to structured problems. More specifically, we will show that an adiabatic search over a subset of the variables can be used to build a better initial Hamiltonian for the global adiabatic search. With this adiabatic algorithm, we recover the same complexity as the nested circuit-based algorithm of Ref.~\cite{cerf00:nesting}.

\section{Adiabatic theorem}
Let us briefly recall the Adiabatic Theorem and how it may be used to design quantum algorithms by adiabatic evolution.

We know that if a quantum system is prepared in the ground state of a time-independent driving Hamiltonian, it remains in this
state. The Adiabatic Theorem states that if this Hamiltonian becomes time-dependent, the system will still stay close to its
instantaneous ground state as long as the variation is \textit{slow enough}.

More specifically, if $|E_0;t\rangle$ and $|E_1;t\rangle$ are the ground and first excited states of the Hamiltonian $H(t)$, with
energies $E_0(t)$ and $E_1(t)$, we define the minimum gap between these eigenvalues
\beq
g_{\min}=\min_{0 \leq t \leq T} \left [ E_1(t)-E_0(t) \right ]
\eeq
and the maximum value of the matrix element of $dH/dt$ between
the eigenstates as
\beq\label{epsilon}
D_{\max}= \max_{0\leq t \leq T}
\left|\langle \frac{dH}{dt} \rangle_{1,0}\right|
\eeq
with $\langle \frac{dH}{dt} \rangle_{1,0}=
\langle E_1;t|\frac {dH}{dt}|E_0;t\rangle$.
The Adiabatic Theorem states that
if we prepare the system at time $t=0$ in its ground state $|E_0;0\rangle$
and let it evolve under the Hamiltonian $H(t)$, then
\beq
|\langle E_0;T| \psi(T) \rangle|^2 \geq 1-\varepsilon^2
\eeq
provided that
\beq\label{adiabatic}
\frac{D_{\max}}{g^2_{\min}} \leq \varepsilon,
\eeq
where $\varepsilon\ll1$.

Now, we may apply to the system a Hamiltonian for which the ground state encodes the unknown solution of a problem.
According to the Adiabatic Theorem, we know that we may get the system very close to this solution state by preparing it in the (known) ground
state of another Hamiltonian, and then by progressively changing it to the Hamiltonian of our problem. This simple idea is central
to the quantum algorithms by adiabatic evolution \cite{farh00,vdam01}.

\section{Quantum search by local adiabatic evolution}\label{qs}
As exposed in \cite{rola02}, this principle may be used to perform a quantum search.
Suppose that among $N$ states, we have to find the $M$-times degenerate ground state of a Hamiltonian
\beq\label{hf}
H_f=I-\sum_{m\in{\mathcal M}}\proj{m},
\eeq
where ${\mathcal M}$ is the ensemble of solutions (of size $M$). We initially prepare the system
in an equal superposition of all could-be solutions:
\beq
|s\rangle=\frac{1}{\sqrt{N}}\sum_{i\in{\mathcal N}}|i\rangle.
\eeq
This superposition is the ground state of the following Hamiltonian:
\beq\label{Hin}
H_i = I-|s\rangle\langle s|.
\eeq
We now apply $H_i$ to the system and switch adiabatically to $H_f$. If we perform an adiabatic evolution
\beq\label{ht}
H(t)=(1-s(t))\,H_i + s(t)\,H_f,
\eeq
where $s(t)$ is a (carefully chosen) monotonic function with $s(0)=0$ and $s(T)=1$,
we will finally obtain a state close to a ground state of $H_f$:
\beq\label{sol}
| \psi_f \rangle\,\approx\,\frac{1}{\sqrt{M}}\sum_{m\in{\mathcal M}}|m\rangle
\eeq
as long as
\beq\label{time}
T=O\left (\sqrt{\frac{N}{M}}\right )
\eeq
This algorithm is refered to as {\em local} because $s(t)$ is chosen such that the Adiabatic Theorem
is obeyed locally, at each time (see \cite{rola02} for details).

\section{Structured problems\label{problem}}
In this article, we consider a class of problems where one has to find an assignment for a set of variables.
For each additional variable considered, new constraints appear and reduce the set of satisfying assignments.
This corresponds to most problems encountered in practice ($k$-SAT, graph coloring, planning, combinatorial optimization, \ldots).

For a set of $n_A$ variables denoted as $A$, there is a corresponding set of constraints $C_A$. We may define a function
$f_A$ that tells if an assignment of the variables in $A$ satisfies the constraints in $C_A$.
\beqa
f_A&:&(\mathbb{Z}_d)^{n_A} \rightarrow \{0,1\}\nonumber\\
&:&x\rightarrow
\left\{\begin{array}{ll}
0 & \hbox{if $x$ does not satisfy $C_A$}\\
1 & \hbox{if $x$ satisfies $C_A$,}
\end{array}\right.\label{function}
\eeqa
where $d$ is the number of possible assignments for each variable ($d=2$ for bits).
As quantum gates have to be reversible, the quantum equivalent of this function will be an oracle:
\beq\label{oracle}
O_A:\mathcal{H}_{N_A}\otimes\mathcal{H}_2\rightarrow\mathcal{H}_{N_A}\otimes\mathcal{H}_2:
\ket{x}\otimes\ket{y}\rightarrow\ket{x}\otimes\ket{y\oplus f_A(x)},
\eeq
where $N_A=d^{n_A}$.
It is shown in Ref.~\cite{rola03} that this oracle is closely related to a Hamiltonian whose ground states, of energy $0$, are the basis states encoding a satisfying assignment and whose excited states, of
energy $1$, are all other basis states:
\beq\label{hamiltonian}
H_A|x\rangle=
\left\{\begin{array}{ll}
|x\rangle & \hbox{if $x$ does not satisfy $C_A$}\\
0 & \hbox{if $x$ satisfies $C_A$}
\end{array}\right.
\eeq
or
\beq
H_A=I_A-\sum_{m_A\in{\mathcal M}_A}|m_A\rangle \langle m_A|,
\eeq
where ${\mathcal M}_A$ is the set of satisfying assignments for the variables in $A$.
It is possible to reproduce the application of this Hamiltonian during a certain time by using
two calls to the quantum oracle $O_A$ (see \cite{rola03} for details).

Now suppose we consider a larger set of variables $n_{AB}=n_A+n_B$ that have to satisfy a set of constraints $C_{AB}\supset C_A$.
To discriminate between assignments satisfying $C_{AB}$ or not, we will use an oracle $O_{AB}$ or a corresponding
Hamiltonian $H_{AB}$ defined as in Eqs. (\ref{oracle}) and (\ref{hamiltonian}). The basic idea of our structured search will be
to find the solutions of $C_{AB}$ by first building the assignments of the $n_A$ primary variables satisfying $C_A$, then
by completing them with all possible assignments of the $n_B$ secondary variables, and finally by searching among these could-be solutions
the global satisfying assignments.

\section{Structured search with two levels of nesting\label{structured}}
This problem is of the same type as the one considered in \cite{cerf00:nesting}, for which the technique of nesting was introduced in the context of the traditional implementation of Grover's algorithm
on a conventional quantum circuit. Here, we apply this technique to the adiabatic quantum search algorithm.

Suppose we divide the variables of our problem into two subsets $A$ ($n_A$ elements) and $B$
($n_B$ elements). First, we will perform a search on the variables in $A$ using the Hamiltonian $H_A$
that encodes the constraints in $C_A$:
\beq
H_A=I_A-\sum_{m_A\in{\mathcal M}_A}|m_A\rangle \langle m_A|.
\eeq
Then we will use the Hamiltonian $H_{AB}$ acting on all variables in $A\cup B$ and encoding the
whole set of constraints $C_{AB}$
\beqa
H_{AB}&=&I_{AB}\\
&&-\sum_{(m_A,m_B)\in{\mathcal M}_{AB}}|m_A\rangle\langle m_A|\otimes|m_B\rangle\langle m_B|\nonumber
\eeqa
to construct a superposition of the solutions of the full problem ${\mathcal M}_{AB}$. A final measurement of the quantum register then gives one of the global solutions at random.

\subsection{Adiabatic search on the primary variables}
The preliminary search on the variables in $A$ is a simple unstructured search as explained in section \ref{qs}. As there are $n_A$ variables in $A$, the corresponding Hilbert space is of dimension
$N_A=d^{n_A}$.
Let $M_A$ be the number of solutions in ${\mathcal M}_A$. Performing an adiabatic quantum
search, we may thus transform the initial state
\beq
|s_A\rangle=\frac{1}{\sqrt{N_A}}\sum_{i\in{\mathcal N}_A} |i\rangle_A
\eeq
into a state close to the uniform superposition of all solutions in ${\mathcal M}_A$
\beq
|\psi_{m_A}\rangle=\frac{1}{\sqrt{M_A}}\sum_{m_A\in{\mathcal M}_A} |m_A\rangle
\eeq
in a time of order
\beq
T_A=0\left (\sqrt{\frac{N_A}{M_A}}\right ).
\eeq

\subsection{Adiabatic search on the secondary variables}
We will now perform a preliminary search in the Hilbert space of dimension $N_B=d^{n_B}$ of the secondary variables in $B$ by extending the partial
solutions $|m_A\rangle$. We prepare the variables in $B$ in a state that is the uniform superposition
\beq
|s_B\rangle=\frac{1}{\sqrt{N_B}}\sum_{j\in{\mathcal N}_B} |j\rangle_B.
\eeq
Globally, the system is thus in the superposition:
\beqa
|\psi_0\rangle_{AB}&=&|\psi_{m_A}\rangle\otimes|s_B\rangle\nonumber\\
&=&\frac{1}{\sqrt{M_A N_B}}\sum_{\newatop{m_A\in{\mathcal M}_A}{j\in{\mathcal N}_B}} |m_A\rangle \otimes |j\rangle_B,\label{psi0}
\eeqa
where some terms correspond to a global solution of the problem [$(m_A,j)\in{\mathcal M}_{AB}$
satisfying all constraints in $C_{AB}$] and the others to a partial solution only [$m_A\in{\mathcal M}_A$
satisfies $C_A$ but $(m_A,j)\notin{\mathcal M}_{AB}$ does not satisfy $C_{AB}$]. We now divide the set
${\mathcal M}_A$ of solutions of $C_A$ into two subsets: ${\mathcal M}_A^S$ will be the set of $m_A$'s
for which there exists at least one solution $(m_A,m_B)$ of $C_{AB}$ and ${\mathcal M}_A^{NS}$ the set
of $m_A$'s for which there is no such solution.
\beqa
{\mathcal M}_A^S&=&\{m_A\in{\mathcal M}_A\,|\,\exists\, m_B, (m_A,m_B)\in{\mathcal M}_{AB}\}\\
{\mathcal M}_A^{NS}&=&\{m_A\in{\mathcal M}_A\,|\,\forall\, j, (m_A,j)\notin{\mathcal M}_{AB}\}
\eeqa
Of course, we thus have ${\mathcal M}_A={\mathcal M}_A^S\cup{\mathcal M}_A^{NS}$. We may now rewrite our
initial state (\ref{psi0}) as
\beqa
|\psi_0\rangle_{AB}&=&\frac{1}{\sqrt{M_A N_B}}\sum_{\newatop{m_A\in{\mathcal M}_A^{NS}}{j\in{\mathcal N}_B}} |m_A\rangle \otimes |j\rangle_B\nonumber\\
&&+\frac{1}{\sqrt{M_A N_B}}\sum_{\newatop{m_A\in{\mathcal M}_A^S}{j\in{\mathcal N}_B}} |m_A\rangle \otimes |j\rangle_B.
\eeqa
In the first part of this expression, no term correspond to a solution of the full problem, whereas in the second part,
some terms do and others do not. The goal of this stage of the computation will be to increase the amplitude of
the solution terms in this last part. To achieve this, we perform an adiabatic evolution using as initial Hamiltonian
\beq\label{h2}
H_i=I_A\otimes(I_B-|s_B\rangle\langle s_B|),
\eeq
that has $|\psi_0\rangle_{AB}$ as a ground state. The final Hamiltonian will be
\beq
H_f=H_{AB}-H_A\otimes I_B.
\eeq
We see that these Hamiltonians share the following properties:
\begin{enumerate}
\item They do not act on states $\ket{i}_A\otimes\ket{s_B}$ corresponding to assignments $i$ of ${\mathcal N}_A$ that do not satisfy $C_A$:\\
$H_{i,f}|i\rangle_A\otimes|s_B\rangle=0 \quad\forall\ i\notin{\mathcal M}_A$.
\item They do not couple states corresponding to different $m_A$'s:\\
$_B\langle j|\otimes\langle m_A|H_{i,f}|m_A'\rangle\otimes|j'\rangle_B=0
\quad\forall\ m_A\neq m_A'\in{\mathcal M}_A,\ \forall\ j,j'\in{\mathcal N}_B$.
\end{enumerate}
It follows that the instantaneous Hamiltonian of the adiabatic evolution $H(t)$ satisfies the same
properties. Keeping this in mind, it may easily be shown that the effect of the adiabatic evolution will be to perform
independent adiabatic searches for each $m_A\in{\mathcal M}_A$. More precisely, each term in $|\psi_0\rangle_{AB}$
\beq
\frac{1}{\sqrt{N_B}}\sum_{j\in{\mathcal N}_B}|m_A\rangle\otimes|j\rangle_B
\eeq
will evolve to
\beq
\frac{1}{\sqrt{M_{B/m_A}}}\sum_{m_B\in{\mathcal M}_{B/m_A}}|m_A\rangle\otimes|m_B\rangle,
\eeq
as long as
\beq
T_{m_A}=O\left (\sqrt{\frac{N_B}{M_{B/m_A}}}\right ),
\eeq
where ${\mathcal M}_{B/m_A}$ is the set of $m_B$'s such that $(m_A,m_B)\in{\mathcal M}_{AB}$ and $M_{B/m_A}$
is the number of these elements.
For this condition to be satisfied for all $m_A$'s simultaneously, we must take
\beq\label{TB}
T_B=\max_{m_A}\, T_{m_A}=O\left(\sqrt{\frac{N_B}{\min_{m_A} M_{B/m_A}}} \right).
\eeq
At the end of this second stage, we thus have constructed a state close to
\beqa
|\psi_{AB}\rangle&=&\frac{1}{\sqrt{M_A N_B}}
 \sum_{\newatop{m_A\in{\mathcal M}_A^{NS}}{j\in{\mathcal N}_B}} |m_A\rangle \otimes |j\rangle_B\nonumber\\
&&+\frac{1}{\sqrt{M_A}} \sum_{m_A\in{\mathcal M}_A^S} e^{i\phi_{m_A}}|m_A\rangle\nonumber\\
&&\otimes \frac{1}{\sqrt{M_{B/m_A}}} \sum_{m_B\in{\mathcal M}_{B/m_A}}|m_B\rangle\nonumber\\
&=&\sqrt{\frac{M_A^{NS}}{M_A}}|\psi^{NS}\rangle+\sqrt{\frac{M_A^S}{M_A}}|\psi^S\rangle,
\eeqa
where $\phi_{m_A}$'s are phases appearing during the evolution,
\beqa
|\psi^{NS}\rangle&=&\frac{1}{\sqrt{M_A^{NS} N_B}}\sum_{\newatop{m_A\in{\mathcal M}_A^{NS}}{j\in{\mathcal N}_B}} |m_A\rangle \otimes |j\rangle_B\\
|\psi^S\rangle&=&\frac{1}{\sqrt{M_A^S}}
 \sum_{m_A\in{\mathcal M}_A^S} e^{i\phi_{m_A}}|m_A\rangle\nonumber\\
&&\otimes\frac{1}{\sqrt{M_{B/m_A}}}\sum_{m_B\in{\mathcal M}_{B/m_A}}|m_B\rangle
\eeqa
and $M_A^{NS}$ (resp. $M_A^S$) is the number of elements in set ${\mathcal M}_A^{NS}$ (resp. ${\mathcal M}_A^S$).

\subsection{Global adiabatic search}
Stages A and B define a unitary evolution $U$ that applies the initial state $|s_A\rangle\otimes|s_B\rangle$
onto $|\psi_{AB}\rangle$:
\beqa
U|s_A\rangle\otimes|s_B\rangle&\approx&|\psi_{AB}\rangle\\
&=&\sqrt{\frac{M_A^{NS}}{M_A}}|\psi^{NS}\rangle+\sqrt{\frac{M_A^S}{M_A}}|\psi^S\rangle.
\eeqa
In this state, we now need to decrease the amplitude of the first term, corresponding to
partial solutions only, and increase the amplitude of the second term, corresponding to global solutions. This could be realized efficiently by performing an adiabatic search using as initial
Hamiltonian:
\beqa
H_i&=&I_{AB}-|\psi_{AB}\rangle\langle\psi_{AB}|\label{h3start}\\
&\approx&U(I_{AB}-|s_A\rangle\langle s_A|\otimes|s_B\rangle\langle s_B|)U^\dagger\\
&\approx&U\,H_0\,U^\dagger,
\eeqa
where $H_0=I_{AB}-|s_A\rangle\langle s_A|\otimes|s_B\rangle\langle s_B|$,
and as final Hamiltonian
\beqa
H_f&=&H_{AB}\label{h3stop}\\
&=&I_{AB}-\sum_{(m_A,m_B)\in{\mathcal M}_{AB}}|m_A\rangle\langle m_A|\otimes|m_B\rangle\langle m_B|\nonumber
\eeqa
during a time
\beq
T_C=0\left (\sqrt{\frac{M_A}{M_A^S}}\right ).
\eeq
Unfortunately, we do not have access to $H_i$, so that the interpolating Hamiltonian $H(s)=(1-s)H_i+s H_f$ cannot be applied directly. However,
we will see in Section~\ref{discretizing} that the basic steps of the quantum circuit implementation of this adiabatic algorithm only require the application of $H_i$ during a particular time $t$, that is
\beq
e^{-iH_it}\approx e^{-iUH_0U^\dagger t}=U e^{-iH_0t} U^\dagger.
\eeq
Hence, each application of $H_i$ during a time $t$ will be equivalent to sequentially applying $U^\dagger$, $e^{-iH_0t}$, and $U$, which
means performing the adiabatic evolution $U$ (stages A and B) backwards, then applying $H_0$ for a time $t$, and finally rerun $U$ forwards (stages A and B).

In section \ref{discretizing}, we will see that, when discretizing the evolution, we must take a number of steps
$r_C$ of order $T_C$. We may now evaluate the complexity of this algorithm. As it consists of $r_C$ steps,
each involving two applications of $U$ or $U^\dagger$, that last a time of order $T_A+T_B$, the
algorithm finally takes a time of order:
\beqa
T&=&(T_A+T_B) r_C\\
&=&O\left(\left(\sqrt{\frac{N_A}{M_A}}+\sqrt{\frac{N_B}{\min_{m_A} M_{B/m_A}}}\right) \sqrt{\frac{M_A}{M_A^S}} \right)\nonumber\\
&=&O\left( \sqrt{\frac{N_A}{M_A^S}}+\sqrt{\frac{M_A N_B}{M_A^S\,\min_{m_A} M_{B/m_A}}}\right).
\eeqa
Let us notice that with the same hypothesis as in Ref.~\cite{cerf00:nesting},
namely
\beq\label{assumption}
M_{B/m_A}=1 \quad \forall\ m_A,
\eeq
then $M_A^S=M_{AB}$, and the computation time is
\beq\label{complexity}
T=O\left( \frac{\sqrt{N_A}+\sqrt{M_A N_B}}{\sqrt{M_{AB}}}\right)
\eeq
i.e., the same complexity as the equivalent circuit-based algorithm exposed 
in Ref.~\cite{cerf00:nesting}. A more detailed analysis of this complexity
will be performed in section \ref{sectioncomplexity}.

\section{Discretizing the adiabatic evolution\label{discretizing}}

\subsection{General method}
The method to implement a \textit{global} adiabatic evolution algorithm on a discrete quantum circuit was initially
exposed in \cite{vdam01}, and was extended to the case of a \textit{local} adiabatic evolution algorithm in \cite{rola03}. Let us quickly review this method, that uses two successive approximations.

The first approximation consists in cutting the evolution time $T$ in $r$ intervals $\Delta T=\frac{T}{r}$ and
replacing the continuously varying Hamiltonian $H(t)$ by a Hamiltonian $H'(t)$ that is constant during each interval $\Delta T$
and varies at times $t_j=j\Delta T$ only.
\beq
H'(t)=H(t_j) \quad\text{if}\quad t_{j-1}\leq t \leq t_j.
\eeq
It is shown in \cite{rola03} that for $H(s)=(1-s)H_i+sH_f$ with $s=s(t)$, this approximation introduces a global error for the evolution
\beq
\opnorm{U(T)-U'(T)}\leq\sqrt{2\frac{T}{r}\opnorm{H_i-H_f}},
\eeq
where $\opnorm{A}=\max_{\|\ket{x}\|=1}\|A\ket{x}\|$ is the operator norm of $A$.
Our algorithm now requires $r$ steps of the form
\beq
U'_j=e^{-iH(t_j)\Delta T}=e^{-i(1-s_j)H_i\Delta T-is_jH_f\Delta T},
\eeq
where $s_j=s(t_j)$.
As we are able to apply $H_i$ and $H_f$ separately but not necessarily a simultaneous combination of them, we will approximate $U'_j$ by
\beq
U''_j=e^{-i(1-s_j)H_i\Delta T}e^{-is_jH_f\Delta T}.
\eeq
This will result in an error
\beq
\opnorm{U'(T)-\prod_j U''_j}\in O(\frac{T^2}{r}\opnorm{[H_i,H_f]})
\eeq
(see \cite{rola03} for details).

\subsection{Application to a structured quantum search}
We now consider the case of a structured quantum search. We could apply the discretization procedure to all three stages (A, B, C) of our algorithm
in order to implement it on a quantum circuit, but we will concentrate on stage C which is the only one that requires discretization.
Nonetheless, it is easy to show that stage A (resp. B) would require a number of steps $r_A$ (resp. $r_B$) of the same order as the computation time
$T_A$ (resp. $T_B$).

For the final stage, the global adiabatic search, the Hamiltonians $H_i$
and $H_f$ are defined in Eq.~(\ref{h3start})-(\ref{h3stop}). Evaluating the errors introduced by the approximations,
we find:
\beqa
\opnorm{H_i-H_f}&<&1\\
\opnorm{[H_i,H_f]}&<&\sqrt{\frac{M_A^S}{M_A}}
\eeqa
and, as $T_C=0\left (\sqrt{\frac{M_A}{M_A^S}}\right )$,
\beqa
\opnorm{U(T)-U'(T)}&\in&O\left(\sqrt{2\frac{\sqrt{\frac{M_A}{M_A^S}}}{r}}\right)\\
\opnorm{U'(T)-\prod_j U''_j}&\in&O\left( \frac{\sqrt{\frac{M_A}{M_A^S}}}{r} \right).
\eeqa
Therefore, as announced in section \ref{structured}, 
we have to cut our evolution in a number of steps
$r_C=O\left (\sqrt{\frac{M_A}{M_A^S}}\right )$ of the same order as $T_C$. Each step $j$ will take the form:
\beqa
U''_j&=&e^{-i(1-s_j)H_i\Delta T}e^{-is_jH_f\Delta T}\\
&\approx&U e^{-i(1-s_j)H_0\Delta T} U^\dagger e^{-is_jH_f\Delta T},
\eeqa
where the applications of Hamiltonians $H_0$ during a time $(1-s_j)\Delta T$ and $H_f$ during a time $s_j\Delta T$ may be realized by
the procedure exposed in \cite{rola03}.

\section{Complexity analysis\label{sectioncomplexity}}
To estimate the efficiency of this algorithm, we will follow the same development as in \cite{cerf00:nesting}:
as we have seen in Section \ref{structured}, under the assumption (\ref{assumption}) that we will consider here, the complexity of this adiabatic algorithm has exactly the same form as its circuit-based counterpart.

First of all let us define a few concepts (for details here and throughout this section, we refer the reader to Ref.~\cite{cerf00:nesting}). The structured search problem is to find an assignment of $n_{AB}=n_A+n_B$ variables among $d$ possibilities and satisfying $e$ constraints, each involving at most $k$ of these variables.
We define as a \textit{ground instance} an assignment of all the variables involved in a particular constraint. A ground instance will be said to be \textit{nogood} if it violates the constraint. Let $\xi$ be the number of those no-good ground instances.

Empirical studies show that the difficulty of solving a structured problem
essentially depends on four parameters: the number of variables $n_{AB}$, the number of possible assignment per variable $d$, the number of variables per constraint $k$, and the number of no-good ground instances $\xi$. Intuitively, we understand that if $\xi$ is small, there are many assignments satisfying the constraints so the problem is easy to solve. On the contrary, if $\xi$ is large, the problem is over-constrained and it is easy to show that there is no solution.
More precisely, it may be shown that for fixed $n_{AB}$ and $d$, the average difficulty may be evaluated by the parameter $\beta=\xi/n_{AB}$.
The hard problems will be concentrated around a critical value $\beta_c$.

Let us now estimate the complexity (\ref{complexity}). Let $p(n)$ be the probability that a randomly generated assignment of the $n$ first variables satisfies all
the constraints involving these variables. We then have $M_A=p(n_A)d^{n_A}$ and $M_{AB}=p(n_{AB})d^{n_{AB}}$ while it is shown in \cite{cerf00:nesting} that:
\beq
p(n)\approx d^{-n_{AB}(\beta/\beta_c)(n/n_{AB})^k}.
\eeq
Eq.~(\ref{complexity}) becomes
\beq
T=O\left(\frac{\sqrt{d^{n_A}}+\sqrt{d^{n_{AB}[1-(\beta/\beta_c)(n_A/n_{AB})^k]}}}{\sqrt{d^{n_{AB}}(1-\beta/\beta_c)}}\right)
\eeq
or, with $a=\sqrt{d^{n_{AB}}}$ and $x=n_A/n_{AB}$:
\beq
T=O\left(\frac{a^x+a^{1-\frac{\beta}{\beta_c}x^k}}{a^{1-\frac{\beta}{\beta_c}}}\right).
\eeq
We now optimize $x$, the fraction of variables for which we perform a partial search, to minimize the computation time.
We have to solve the equation
\beq
\frac{\beta}{\beta_c}kx^{k-1}=a^{\frac{\beta}{\beta_c}x^k+x-1}
\eeq
which, for large $a$ (that is large $n_{AB}$) approximately reduces to
\beq\label{alpha}
\frac{\beta}{\beta_c}x^k+x-1=0.
\eeq
The solution of this equation $\alpha$ ($0\leq\alpha\leq1$) corresponds to the optimal partial search we may perform such that the complexity grows with the smallest power in $d$ for $n_{AB}\to\infty$. This optimal computation time may then be written as
\beq
T=O\left(\frac{2a^\alpha}{a^{1-\frac{\beta}{\beta_c}}}\right)=O\left( \frac{\sqrt{d^{\alpha n_{AB}}}}{\sqrt{d^{n_{AB}(1-\frac{\beta}{\beta_c})}}} \right).
\eeq
Let us now consider the hardest problems for which $\beta\approx\beta_c$. For these problems, the complexity reads:
\beq
T=O\left( \sqrt{d^{\alpha n_{AB}}}\right)
\eeq
that we may immediately compare to the complexity of an unstructured quantum search $O\left( \sqrt{d^{n_{AB}}}\right)$. The gain in the exponent $\alpha$ depends on
$k$ through Eq.~(\ref{alpha}) For instance, we find $\alpha=0.62$ for $k=2$, $\alpha=0.68$ for $k=3$ and $\alpha\to1$ when $k\to\infty$.

To compare these results with a classical algorithm, let us consider a specific problem, the satisfiability of boolean formulas in conjunctive normal form, or $k$-SAT.
For $3$-SAT, which is known to be NP-complete, the best known classical algorithm has a worst-case running time that scales as $O\left( 2^{0.45n_{AB}} \right)$ \cite{paturi},
while, as $\alpha=0.68$ for $k=3$, our quantum adiabatic algorithm has a computation time of order $O\left( 2^{0.34n_{AB}} \right)$, which is a slight improvement.
Let us also notice that this scaling could be further improved by using several levels of nesting, i.e. by replacing the preliminary search over the primary
variables by a another nested structured search (see the analysis of the circuit-based counterpart of this idea in the Appendix of \cite{cerf00:nesting}).

\section{Conclusion}
We have introduced a new quantum search algorithm combining the approach based on local adiabatic evolution developed in \cite{rola02} and the nesting technique introduced in \cite{cerf00:nesting}. It allows one to adiabatically solve structured search problems with an improved complexity over a naive adiabatic search that would not exploit the structure of the problem.

The basic idea is to perform a preliminary adiabatic search over a reduced number of variables of the problem in order to keep only a superposition of the assignments that respect the constraints of this partial problem, and then to complete these partial solutions by finding satisfying assignments for the remaining variables. We have seen that to implement this algorithm, the global adiabatic evolution (stage C) has to be discretized, which makes it possible to \textit{nest} the preliminary search (stages A and B) into the global one.
Each step of the algorithm requires to alternate partial adiabatic searches backwards and forwards with global search operations.

A complexity analysis shows that the average computation time of this adiabatic algorithm, although still exponential, 
grows with a reduced exponent compared to quantum
unstructured search algorithms or classical known algorithms to solve a problem such as $3$-SAT.

\begin{acknowledgments}
J.R. acknowledges support from the Belgian foundation FRIA.
This work was funded in part by the Communaut\'e Fran\c caise de
Belgique under grant ARC 00/05-251, by the IUAP programme of the Belgian
government under grant V-18, and by the EU under project RESQ
(IST-2001-35759).
\end{acknowledgments}

\bibliography{qit,nesting}

\end{document}